\PassOptionsToPackage{colorlinks=true}{hyperref}
\documentclass[sigconf,10pt]{acmart}

\renewcommand\footnotetextcopyrightpermission[1]{}
\setcopyright{none}

\usepackage[english]{babel}

\usepackage{enumitem}
\setitemize{leftmargin=4mm}
\setenumerate{leftmargin=4mm}
\setlist{noitemsep,topsep=0pt,parsep=0pt,partopsep=0pt}

\newcommand{\paraskip}{\vspace{2pt}}
\newcommand{\parahead}[1]{\paraskip\noindent\textbf{#1}}

\usepackage{cleveref}
\crefname{figure}{Fig.}{Figs.}
\crefname{section}{\S}{\S}
\crefname{table}{Tab.}{Tab.}
\crefname{algorithm}{Alg.}{Algs.}
\crefname{equation}{Eq.}{Eqs.}
\crefrangelabelformat{section}{#3#1#4-#5#2#6}
\crefrangelabelformat{figure}{#3#1#4-#5#2#6}
\usepackage{amsmath}
\usepackage[hang,flushmargin]{footmisc}
\usepackage{balance}
\usepackage{subcaption}
\usepackage[utf8]{inputenc}
\usepackage{float}
\usepackage{totpages}
\usepackage{booktabs}

\usepackage{wici-techreport}
\ReportLogoPath{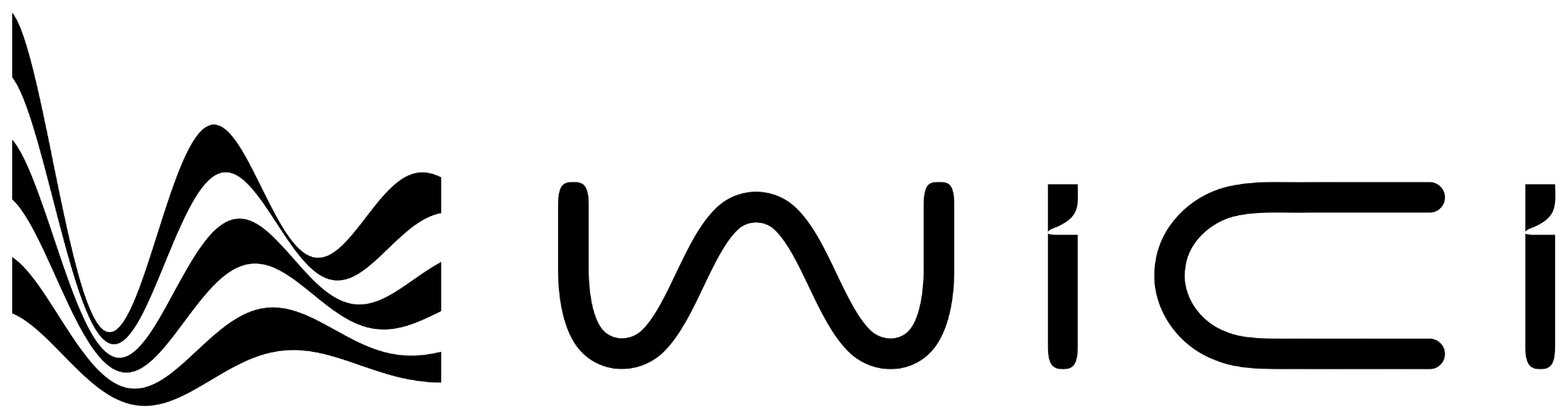}
\ReportTitleLogoYOffset{1.6mm}
\ReportKicker{WiCi TECHNICAL REPORT}
\ReportNumber{WiCi-TR-2026-001}
\ReportDate{August 2026}
\ReportURL{}
\ReportHeaderTitle{WiCi}

\acmDOI{}
\acmISBN{}
\acmPrice{}

\begin{document}

\title{WiCi: Wireless GPU Computing Infrastructure}

\author{Yibin Shen*, Wei Li*, Kaiqiang Xu, Zili Meng\\hello@wici.ai}
\renewcommand{\shortauthors}{WiCi Technical Report}

\begin{abstract}
LLM inference applications are gaining significant traction.
The demand for inference is growing exponentially, and the GPU usage of inference is increasingly surpassing that of training.
Due to the mobility penalty, edge-side inference fails to deliver satisfactory performance.
Consequently, most inference service providers currently rely on cloud-based inference, which incurs substantial, not sustainable costs for enterprises, and is even increasing in the agentic paradigm.
Therefore, our goal is to enable powerful computing capabilities as server-grade GPUs on mobile devices.
We propose Wireless GPU Computing Infrastructure (WiCi) in this paper.
Through WiCi, mobile devices can wirelessly access server-grade GPUs, running inference tasks on mobile clients but offloading GPU-related computations to a nearby GPU via WiFi.
WiCi introduces a series of designs to make sure the infrastructure is scalable with different applications, compatible with different mobile devices, and has comparable performance to running on a physical GPU.
We test WiCi from mobile devices and find that WiCi can reduce time to first token by up to 90\%, improve the token rate by approximately 39$\times$ compared to local inference on mobile devices for the same model, and support much larger models.
WiCi also achieves up to nearly 80\% of the native performance of the server-grade GPU across different applications.
\end{abstract}

\maketitle
\hypersetup{%
  pdftitle={WiCi: Wireless GPU Computing Infrastructure},
  pdfauthor={Yibin Shen*, Wei Li*, Kaiqiang Xu, Zili Meng\\hello@wici.ai},
}

\let\thefootnote\relax\footnote{*Equal contribution.}
\section{Introduction}
The rapid adoption of large language models (LLMs) in daily applications has led to an explosive increase in inference workloads. 
For example, it is estimated that Google's AI systems consume 1.3 quadrillion ($10^{15}$) tokens per month, and this figure continues to grow as usage expands. 
The shift toward agent applications further amplifies the inference and communication overhead, since: 
(i) A single user task now requires multiple iterative LLM interactions, and 
(ii) Some agentic applications need to continuously stream the user contents (e.g., a screenshot) to the LLM. 
This further magnifies overall inference and communication demand, making total token consumption grow superlinearly.



There are two typical ways of serving LLM tasks to users: cloud solution and mobile solution.
The cloud solution involves service providers deploying large-scale GPU clusters to handle user requests, which are commonly used in state-of-the-art commercial services like GPT \cite{gpt} and Gemini \cite{gemini}.
However, it imposes high costs on operators for inference, requiring substantial investments.
For example, OpenAI is projected to spend USD 1 trillion on computing infrastructure in 2025-2035~\cite{openaicost}, where more than 73\% goes to inference rather than training~\cite{deloitte2025inferenceratio}.
The mobile solution entails running the application entirely on the user's mobile device (smartphones and laptops) using the device's GPU.
However, due to the huge performance gap between mobile GPUs and server-grade GPUs (can be 100x, \cref{sec:motivation_serving_llm}), only smaller edge LLMs are supported.

Our primary motivation is to enable powerful computing capabilities at mobile devices without relying on cloud services. 
A straightforward idea is to put the server-grade GPUs onto mobile clients; however, this solution can be challenging due to weight and power constraints. Standalone GPUs significantly compromise the mobility of mobile devices. For instance, smartphones typically weigh only 200 grams and have a power of several watts, whereas server GPUs, such as the NVIDIA RTX 5090, weigh 2 to 3 kilograms and require up to 600 watts of power supply, both of which are still increasing.


\begin{figure}
    \centering
    \captionsetup{font=small}
    \includegraphics[width=0.9\linewidth]{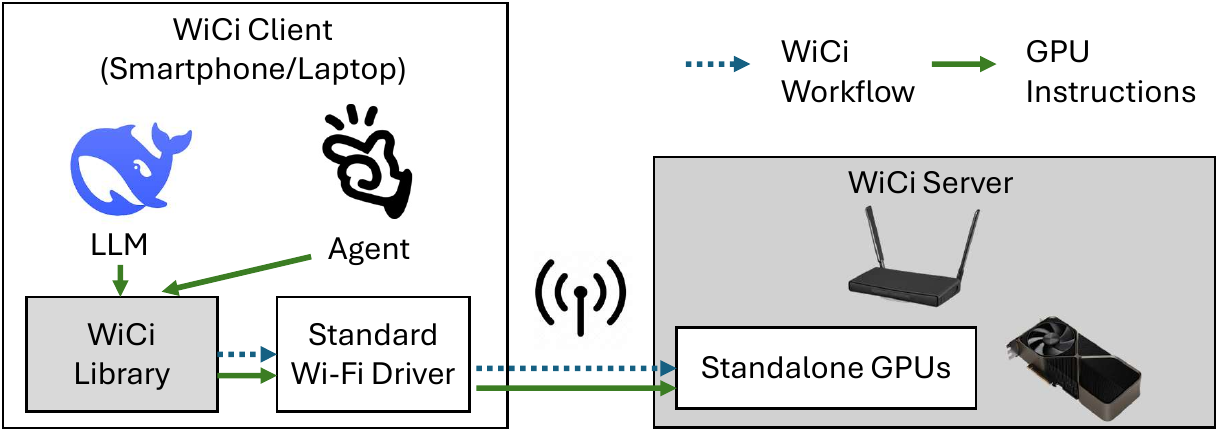}
    \setlength{\abovecaptionskip}{1mm}
    \caption{WiCi overview.}
    \vspace{-7mm}
    \label{fig:intro}
\end{figure}

So here we come up with our design — Wireless GPU Computing Infrastructure (WiCi).
To summarize the desired outcome in one sentence: wirelessly access a GPU without adding any hardware, as if a truly zero-weight GPU is directly connected to the mobile devices as shown in \cref{fig:intro}.
In other words, we aim at running inference tasks locally on mobile devices, but offloading GPU-related computations to a nearby GPU via wireless connection (e.g., WiFi). 
In this way, we can have comparable performance of cloud solutions but enjoy the mobility for mobile clients.
Agentic applications no longer need to frequently stream the contents to the cloud, and LLM service providers can significantly save the operational expenses.


Nevertheless, we have three challenging design goals:
\begin{itemize}
\item \textbf{Scalable to applications.} We aim to deliver a seamless transition for inference applications from local computing to our architecture without requiring any modifications at the application layer, providing an app-agnostic experience. 
\item \textbf{Compatible to devices.} We need to be compatible with various devices and operating systems. Therefore, we should not intrusively modify the mobile clients with features that are not supported by the commercial firmware.
\item \textbf{Comparable performance.} We need to overcome the performance gap between PCIe and the wireless link. GPU computing tasks involve massive data volumes and frequent interactions which are typically through PCIe. WiFi measures in milliseconds while PCIe is microsecond-level, and WiFi delivers data at up to 1 Gbps while PCIe achieves tens of Gbps. Mitigating its impact is our primary focus.
\end{itemize}
Existing solutions cannot satisfy three goals at the same time. 
Existing approaches for remote GPU access generally fall into two categories: application-level partitioning or low-level virtualization. 
Application-level offloading (e.g., remote PyTorch) requires invasive modifications to source code, resulting in huge case-by-case engineering efforts, which are not scalable to applications. 
Low-level techniques, such as PCIe forwarding or MMIO interception, need to invasively modify the mobile client side, which is not feasible for large-scale usage.  
Even this, if we simply connect to a remote GPU server using existing API translation frameworks like rCUDA~\cite{duato2010rcuda} and sCUDA~\cite{scuda} through TCP/IP, the performance will still be drastically degraded. It is non-trivial to achieve a comparable performance.

Our solution is to place a GPU at a WiFi router that originally supports WiFi connectivity and a stable power supply.
WiCi intervenes at the user space driver level to ensure the scalability to applications and compatibility with devices.
We intercept GPU computing tasks below the user-space program and above the kernel-mode driver, forwarding them to the GPU router for remote execution. 
Applications perceive GPU computations as completed locally on the mobile device.
As a result, we are no longer constrained by the hardware capabilities of mobile devices, enabling us to enjoy performance comparable to server GPUs. 

To enhance performance, we systematically propose a series of designs tailored to the function call characteristics of inference applications.
First, to minimize the data and model weight transmission between CPU and GPU through WiFi due to its limited bandwidth, we introduce a caching mechanism at the router side.
Second, to avoid the frequent blocking due to interaction delay in the sequential execution of GPU instructions, we introduce speculative execution to group multiple functions in one batch.
Third, to reduce the transmission of intermediate results and extensive interactions, we exploit the similarity and cyclical nature of functions called when LLM generates tokens, and introduce a tracing mechanism to replay the function traces for multi-layer inference.


We deploy the WiCi infrastructure on a NVIDIA 4090D 48G GPU equipped with WiFi 6 and use a Raspberry Pi 5 16GB as the mobile device.
We conducted comprehensive evaluations across various application types. When applied to different LLM applications using llama.cpp, WiCi can achieve up to 65 — 80\% of the native performance of the standalone GPU, resulting in 1-2 orders of magnitude improvements compared to running with mobile computing capability.
Additionally, we have demonstrated WiCi's robustness in handling diverse network conditions and managing multiple client requests simultaneously.

Our main contributions can be summarized as follows:
\begin{itemize}
    \item We implement WiCi to provide application-scalable, device-compatible, and high-performance GPU computing resources to mobile devices.
    \item We propose various designs, including model caching, batched execution, and trace replay, to minimize the performance degradation brought with wireless networking.
    \item We deploy the WiCi system on a local testbed and conduct comprehensive evaluations, including various application types and the impact of variations in different metrics.
\end{itemize}

\section{Background and Motivations}
In this section, we will introduce the exponential rise in inference demand (§\ref{sec:motivation_exponential}), the current state of serving LLMs (§\ref{sec:motivation_serving_llm}),  the inference capability requirement (§\ref{sec:motivation_inference_capability}), the design goals (§\ref{sec:motivation_motivations}), as well as the discussion about existing work (§\ref{sec:motivation_existing_work}), driving us to carry out this new infrastructure WiCi.

\begin{figure}[t]
     \centering
     \captionsetup{font=small}
     \begin{subfigure}[t]{0.48\linewidth}
            \centering
            \setlength{\abovecaptionskip}{0mm}
            \includegraphics[width=\textwidth]{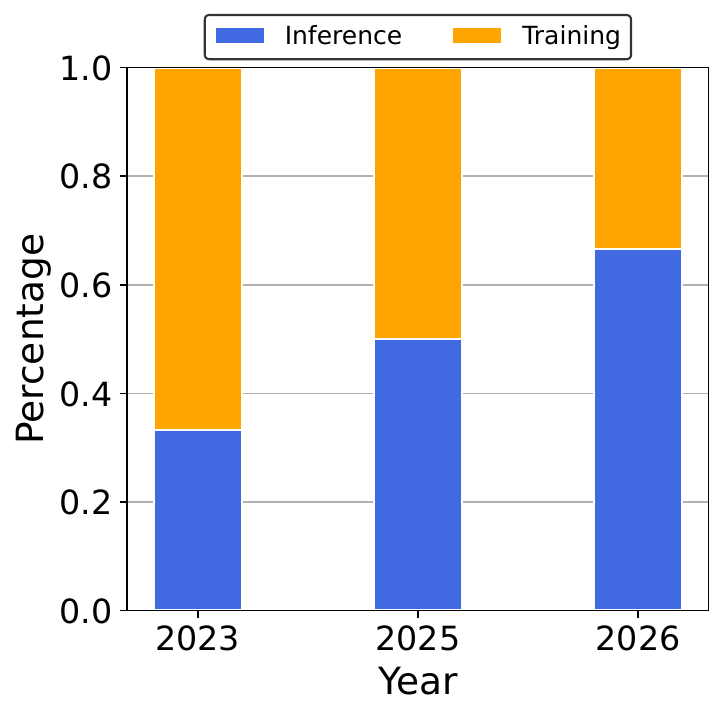}  
            \caption{Inference Cost Ratio}
            \label{fig:motivation_ratio}  
     \end{subfigure}
     \hspace{1mm}
     \begin{subfigure}[t]{0.48\linewidth}
            \centering
            \setlength{\abovecaptionskip}{0mm}
            \includegraphics[width=\textwidth]{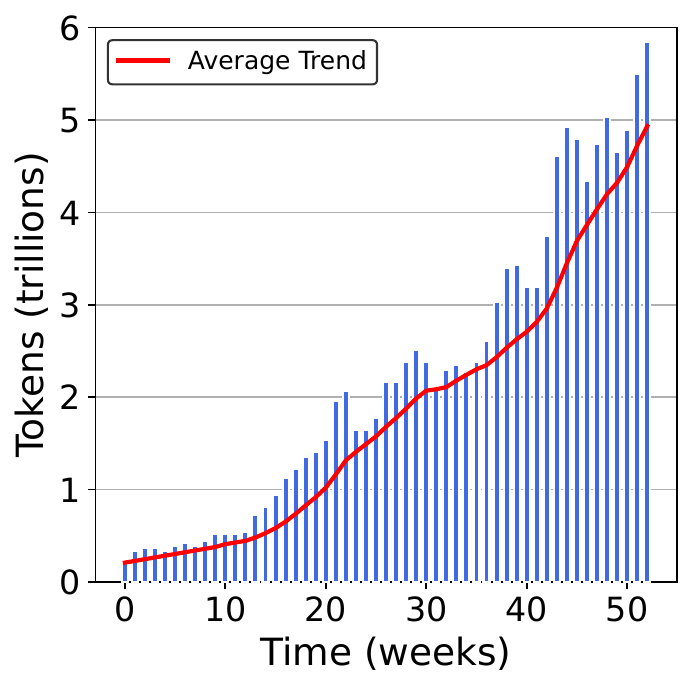}  
            \caption{Weekly Token Consumption}
            \label{fig:motivation_token_consume}  
     \end{subfigure}
     \setlength{\abovecaptionskip}{0mm}
     \caption{(a) The proportion of inference costs increases from 2023 to 2026. (b) The weekly tokens consumed by main LLMs grow exponentially from 11/04/2024 to 11/03/2025.}
     \vspace{-6mm}
\end{figure}

\subsection{Exponential Rise in Inference Demand}
\label{sec:motivation_exponential}
\parahead{Increasing inference proportion and demands.} 
As the number of users accessing LLM services continues to increase daily, the inference costs for service providers are surpassing the costs from training.
For example, the inference workloads have been increasing in the past years, from 33\% (2023) to 50\% (2025), and to 70\% (2026), as shown in \cref{fig:motivation_ratio} \cite{deloitte2025inferenceratio}. 
In the meantime, the total tokens have also been increasing (\cref{fig:motivation_token_consume})  \cite{a16z2025tokenconsumption}. 
In one year, weekly token usage increased from 0.3 trillion to 6 trillion, demonstrating an exponential growth trend. 

\parahead{Agentic paradigm further contributes to the exponential increase in inference demand. }
The concept of AI agents has recently exploded in popularity, yet their inference modes are far more complex than simple text-based Q\&A or speech recognition. 
Take Doubao mobile phone~\cite{doubao2026aiphone} and Manus as examples.
Doubao mobile phone's GUI agent accepts voice and screen recording inputs and then schedules cloud-mobile inference.
In this procedure, frequent interactions between the LLM and the mobile phone are required.
Similarly, Manus can consume one million tokens to generate a location-based daily web application since it encompasses various subtasks, including data collection, data visualization, and programming. 
Thus, the token consumption differs in agentic applications significantly from conventional AI computing modes. And the annual token consumption surges from 0.0005 PetaTokens (2025) to 152,667 PetaTokens (2030), with an annual growth rate of 3418\% \cite{idc2025agentconsumption}.

\parahead{The explosion in inference demand brings high costs to service providers. }
In response to this increasing demand, numerous cloud computing providers are making significant investments in enhancing their cloud infrastructure. 
For instance, Anthropic allocates between 50\% and 65\% of its revenue to cover inference expenses. 
Similarly, OpenAI's inference costs are reported to be three times greater than its personnel expenses. 
%
These substantial costs will ultimately be passed on to users, although many services are free now. 
As one extreme example, ChatGPT Pro subscription costs \$199 per month while NVIDIA RTX 5090 sells at \$1,999, which only costs 10 months of ChatGPT Pro.
Under such circumstances, the current approach of relying on cloud computing power for inference appears not to be financially sustainable.

\begin{figure}[t]
     \centering
     \captionsetup{font=small}
     \begin{subfigure}[t]{0.5\linewidth}
            \centering
            \setlength{\abovecaptionskip}{0mm}
            \includegraphics[width=\textwidth]{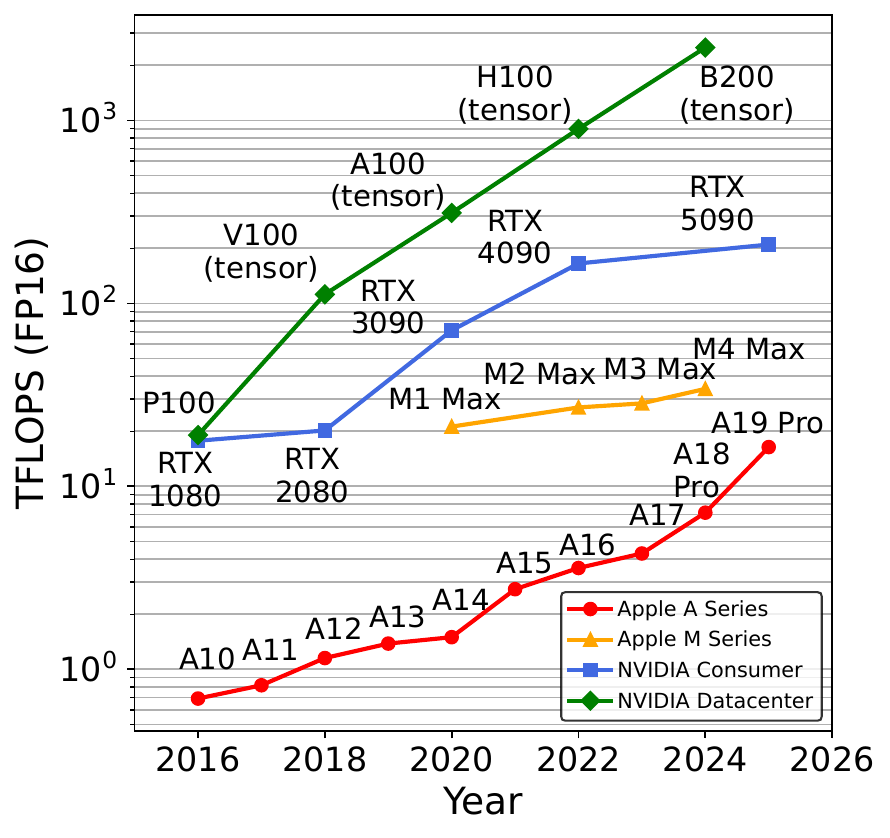}
            \caption{TFLOPS Trend}
            \label{fig:motivation_GPU_year_flops}  
     \end{subfigure}
     \hspace{-1mm}
     \begin{subfigure}[t]{0.48\linewidth}
            \centering
            \setlength{\abovecaptionskip}{0mm}
            \includegraphics[width=\textwidth]{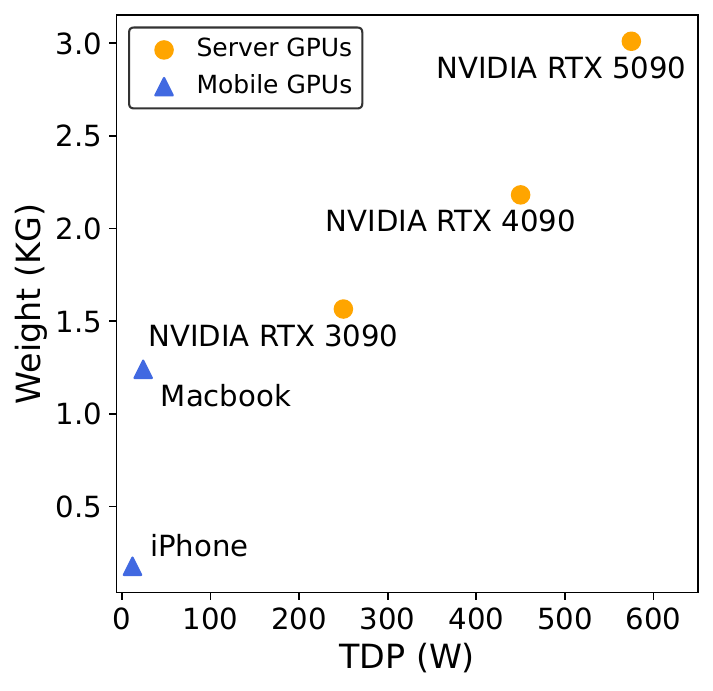}  
            \caption{TDP vs Weight}
            \label{fig:motivation_server_mobile_gpu}  
     \end{subfigure}
     \setlength{\abovecaptionskip}{0mm}
     \caption{(a) The gap in computation capability between mobile GPUs and server-grade GPUs does not narrow from 2016 to 2026. (b) The thermal design power (TDP) of server GPUs is much higher, and the weight of server GPUs is much larger compared to mobile GPUs.}
     \vspace{-6mm}
\end{figure}


\subsection{Serving LLMs}
\label{sec:motivation_serving_llm}
There are two typical ways of serving LLMs – the cloud solution and the mobile solution.
\begin{itemize}
    \item Cloud solution: the service providers deploy large-scale GPU clusters to handle the requests from the users, and users send the requests to the cloud for the responses. This has already been widely used for state-of-the-art commercial services such as GPT or Gemini. 
    \item Mobile solution: the application runs completely in the client’s mobile device (mostly smartphones and laptops) and utilizes the computing capabilities of the mobile GPU. The LLM here has to be smaller and underperform.
\end{itemize}
However, both solutions have their limitations.
Cloud solutions result in a significant burden for the operators on both the inference costs and communication overheads. To process the quadrillion-scale tokens, major providers have therefore committed extraordinary investments – e.g., OpenAI is projected to spend USD 1 trillion on computing infrastructure in 2025-2035~\cite{openaicost}. It is worth noting that while training is a one‑off expense, inference cost persists and scales with user demand. As a concrete example, over 73\% GPU usage is attributed to inference rather than training nowadays, and the proportion is still increasing. Meanwhile, the frequent communications from agentic applications are also limited by the cloud interaction.

Mobile solutions, on the other hand, suffer from the mobility penalty since there is a tradeoff between performance and mobility. Mobility penalty refers to the significant gap in computing power between lightweight mobile devices and servers, primarily due to constraints such as power supply and weight limitations. For mobile processors that fit mobile clients (e.g., Qualcomm Snapdragon), the computing capability is way behind that of server-grade GPUs. For example, as shown in \cref{fig:motivation_GPU_year_flops}, NVIDIA series GPU is always 1-2 orders of magnitudes better than that of iPhone's processing units (A series).



We further examine the trend of differences between mobile GPUs and server-grade GPUs, and find that the gap is also increasing.
We present the computing capability of four series of chips: Apple A series (used in iPhones), Apple M series (used in MacBooks), NVIDIA consumer GPUs, and NVIDIA datacenter GPUs, as shown in Fig. \ref{fig:motivation_GPU_year_flops}. 
From 2016 to 2026, all four chip series have increased; however, the gap in computational capability between smartphone chips and datacenter GPUs even goes up.
Specifically, in 2016, NVIDIA P100 is 30x that of Apple A10, while the gap in 2024 between B200 and A18 Pro is 350x.

\noindent\textbf{Mobility penalty. }
The performance gap comes from the weight and power limitations of mobile devices.
%
%
Mobility penalty refers to the significant gap in computing power between lightweight mobile devices and servers, primarily due to constraints such as power supply and weight limitations.
As shown in \cref{fig:motivation_server_mobile_gpu}, the design power of MacBook and iPhone is several orders of magnitudes lower than that of NVIDIA consumer GPUs, let alone datacenter GPUs.
In the meantime, the weight of smartphones and common laptops is also much lower than that of server-grade GPUs.
Even though NVIDIA is actively developing mobile GPUs, the gap is estimated to remain around a hundredfold by 2025.

Therefore, we plan to propose a new computing infrastructure that can directly bridge this performance gap.
We intend to replace traditional hardware connections between GPUs and hosts with WiFi. 
This approach leverages the inherent wireless connectivity of mobile devices, enabling server-grade GPUs to connect weightlessly to mobile devices.

\begin{figure}
    \centering
    \captionsetup{font=small}
    \includegraphics[width=0.85\linewidth]{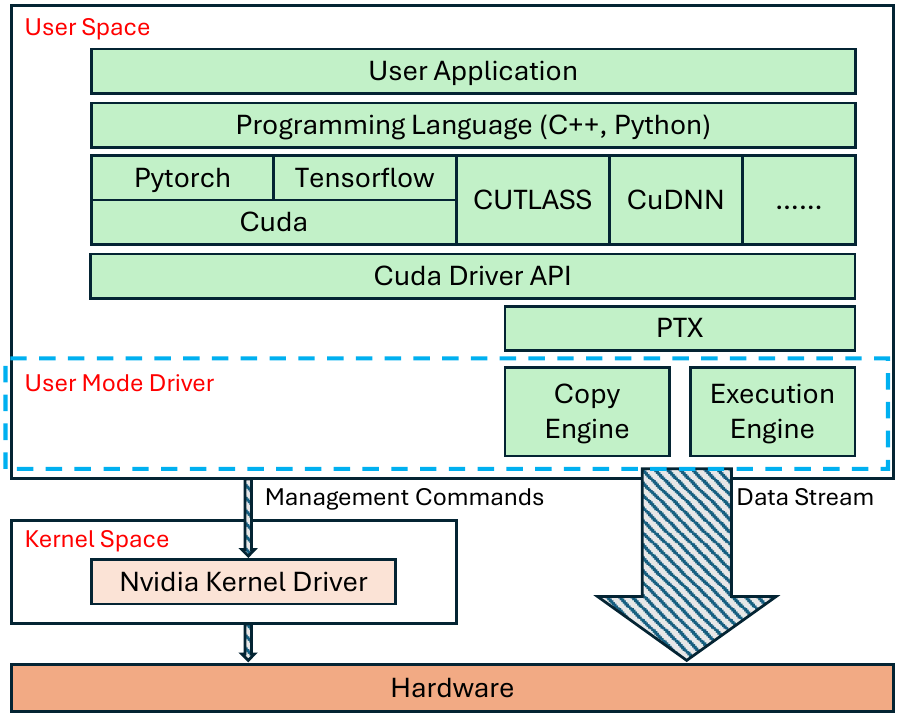}
    \setlength{\abovecaptionskip}{2mm}
    \caption{NVIDIA GPU software stack.}
    \vspace{-6mm}
    \label{fig:cloud_infra}
\end{figure}

\subsection{Inference Capability Requirement}
\label{sec:motivation_inference_capability}
Our observation here is that while training LLM typically requires tens of thousands of GPUs, serving the LLM for inference does not require that many GPUs.
In fact, many leading LLMs fit in the size of one desktop GPU.
We explain the requirement from two perspectives -- memory and inference speed.

For inference speed, if the model can fit into one GPU, the inference speed will have no difference no matter the GPU is in the cloud or with WiCi.
For example, Qwen3-32B \cite{yang2025qwen3} demonstrates performance comparable to Gemini 2.5-Pro, but the model size of Qwen3-32B is 65.5 GB. 
In this scenario, according to the benchmarking results provided in \cite{qwendocs}, deploying an NVIDIA H20 with 96 GB of memory can achieve inference speeds of hundreds of tokens per second, which is deemed adequate for user applications.

Some LLMs might be too large to fit into one GPU, while the design of WiCi is still scalable to such options.
For example, some LLMs containing 200 to 300 billion parameters, which necessitate 500 to 600 GB of GPU memory. 
Even in this case, we can configure WiCi to be equipped with eight NVIDIA H20 96GB GPUs, or more advanced models like B200, enabling inference speeds of hundreds of tokens per second.
In this case, mobile phones can still wirelessly enjoy the high-performance GPU as long as there is a need.
Later in this paper, we demonstrate that one standalone server-grade GPU has already been sufficient for major models.




\subsection{Design Goals}
\label{sec:motivation_motivations}

Our primary motivation is to address the challenge of achieving high computational capabilities and mobility without compromise on mobile devices by leveraging a new computing infrastructure. And the goal is to achieve three key performance objectives:

\begin{itemize}
    \item \textbf{Scalable to applications.} We aim to deliver a seamless transition for inference applications from existing local computing logic to our architecture without requiring any modifications at the application layer, providing an app-agnostic experience. 
    \item \textbf{Compatible to devices.} We need to be compatible with various devices and operating systems. Therefore, we should not intrusively modify the mobile clients with features that are not supported by the existing commercial hardware and firmware.
    \item \textbf{Comparable performance.} We need to overcome the performance gap between PCIe and the wireless link. GPU computing tasks involve massive data volumes and frequent interactions. WiFi RTT typically measures in milliseconds while PCIe is microsecond-level, and WiFi delivers data at up to 1 Gbps while PCIe achieves tens of Gbps. Mitigating its impact is our primary focus.

\end{itemize}

\subsection{Why Existing Solutions Fail?}
\label{sec:motivation_existing_work}
Here we explain the other solutions for remotely connecting to a GPU and why they cannot meet all design goals above.

We use the NVIDIA GPU as an example due to its wide adoption.
\cref{fig:cloud_infra} depicts the complete NVIDIA software stack from application to hardware. 
NVIDIA provides CUDA libraries for various programming languages, which allow developers to utilize GPU power for scientific computing and machine learning. 
It consists of two main parts: the CUDA runtime and driver API, which are largely interchangeable but differ in complexity and context management. 
Additionally, NVIDIA offers parallel thread execution (PTX), a low-level language that serves as an intermediate representation in the CUDA model, enhancing optimization and portability across GPU architectures. LLM inference frameworks, such as llama.cpp \cite{llamacpp}, vLLM \cite{kwon2023efficient}, and SGLang \cite{zheng2024sglang}, utilize CUDA and PTX for optimal performance on NVIDIA GPUs. 
All user applications and libraries must allocate GPU resources through the kernel driver.

Therefore, when utilizing a remote GPU to perform a task, it is possible to intercept and forward any layer of this stack to a remote host for execution, a technique commonly referred to as API interception. 
Subsequently, we will analyze the shortcomings of existing solutions at each layer of the stack and justify the design choice of WiCi.

\textit{High level: Application level interception is not scalable. } 
The first option is to intercept at the application level, e.g., programming languages, and libraries such as TensorFlow \cite{abadi2016tensorflow}, PyTorch \cite{paszke2019pytorch}, vLLM \cite{kwon2023efficient}, and CuDNN. 
Consequently, we can explore the interception of the Python frameworks to achieve specific objectives, as demonstrated by AntMan \cite{xiao2020antman}. 
However, a significant limitation arises in terms of scalability to applications. 
On one side, it requires to intrusively modify every library given the diversity of existing programming languages and libraries used by LLM developers.
On the other side, when aiming for optimal performance, state-of-the-art frameworks might directly use C++ or even assembly language rather than high-level modules. 
In such cases, this solution may be unable to intercept the necessary functions effectively.

\textit{Median level: adequate scalability but requires performance optimization.}
By intercepting at the level of CUDA and PTX code, the compatibility of the solution can be significantly improved, irrespective of the programming language used. 
Existing solutions such as rCUDA \cite{duato2010rcuda}, sCUDA \cite{scuda}, and ZLUDA \cite{zluda} intercept at the CUDA user libraries, and KRYPTON~\cite{zhang2025efficient} intercepts at the CUDA kernel level.
However, they suffer from severe performance degradation when over the wireless network. 
This is because they are architected for stable, high-bandwidth wired interconnects for the purpose of virtualization or resource isolation, and lack the latency-hiding mechanisms necessary to handle the stochastic jitter of wireless links. 
We later demonstrate in the evaluation (\cref{sec:batching_trace} and \cref{sec:eval}) of the performance degradation.

\textit{Low level: PCIe interception is not device-compatible, and further degrades performance. } 
At the lowest level of the technology stack, any modification will require the hardware or operating system to be intrusively changed, which is not feasible for most of mobile clients (e.g., smartphones or laptops).
Such a design will drastically limit the deployability since this requires collaboration with manufacturers.
Meanwhile, directly intercepting PCIe instructions will expose the performance gap between WiFi and PCIe directly.
Therefore, it is impractical to intercept at the low level in the stack.

Overall, developing at the median level is the most suitable choice as it strikes a balance between compatibility and performance.
Inside this level, we further choose the user-mode driver as the most suitable component to modify.
If we make modifications at lower parts of the median level, that is, in NVIDIA kernel drivers or PTX, we will encounter two issues.
\begin{itemize}
    \item Limited scalability. The primary reason for limited scalability is that NVIDIA's kernel drivers themselves impose certain platform and operating system constraints. Currently, NVIDIA's open-source kernel drivers only support select Linux systems. Currently, NVIDIA's open-source driver only supports select Linux systems. Additionally, for function calls from higher-level to be properly handed off to the kernel driver for execution, the native system of the mobile devices must fully support CUDA libraries and the CUDA user-mode driver. This is simply not feasible on many mobile devices without GPUs.
    \item Largely degraded performance. As shown in \cref{fig:rtt_analysis_chart}, we measured the job completion time (JCT) for the most basic "nvidia-smi" command. When executed locally, the average completion time is only 0.051 seconds. However, when we implement modifications in the kernel driver, hijacking and forwarding functions to the remote GPU within the kernel driver, the time cost increases by $51\times$. The primary reason is that completing the same task requires more function calls at lower levels. Each function call incurs at least one round-trip time (RTT) for remote execution. The more function calls involved, the worse the performance becomes. In kernel drivers, remotely executing "nvidia-smi" consumes nearly 800 RTTs, resulting in unacceptable performance degradation.
\end{itemize}

\section{WiCi Design}

\subsection{Basic Idea and Framework Overview}
\begin{figure}
    \centering
    \captionsetup{font=small}
    \includegraphics[width=\linewidth]{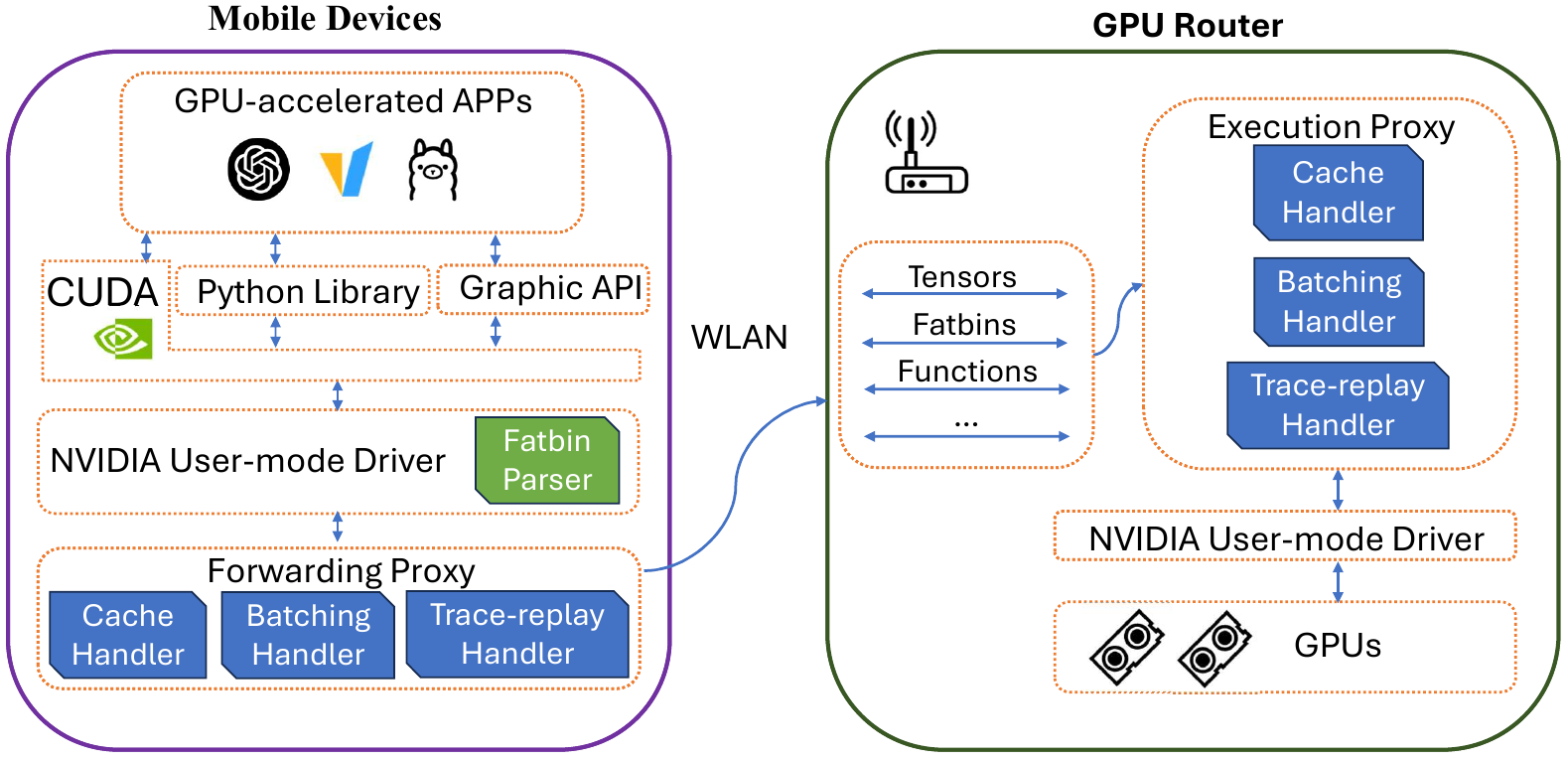}
    \setlength{\abovecaptionskip}{-3mm}
    \caption{Framework of WiCi.}
    \vspace{-6mm}
    \label{framework}
\end{figure}


The basic idea for our proposed infrastructure is to leverage WiFi in a manner similar to PCIe, enabling GPUs to connect to mobile devices via WiFi. 
In other words, unlike mobile or cloud inference solutions, we keep input preprocessing on the mobile device, and also retain part of the NVIDIA software stack on mobile devices.
We offload only the components that must run on the GPU to the GPU router, specifically the user-mode driver and its underlying layers.
This approach allows devices to retain mobile inference experience while benefiting from server-level GPU capabilities delivered over WLAN.

As shown in \cref{framework}, WiCi's core design takes place within NVIDIA's user-mode driver. 
When upper-layer software libraries invoke the CUDA driver, we intercept these calls within the driver itself. 
Functions that would normally be passed to the kernel-space driver, along with data communications with the GPU firmware, are instead redirected to the remote GPU.
We chose to implement modifications at this layer because it allows us to maximize scalability while ensuring an app-agnostic experience for the users.

\vspace{-0.4cm}
\subsection{WiCi Library}
After selecting the level at which we develop our forwarding strategy, we further give the design of the WiCi library, which is the core part of API forwarding.

\noindent\textbf{API forwarding.}
CUDA is not yet a fully open standard, and some internal details have not been officially documented; only the APIs are publicly available. Thus, the only way to do API forwarding is to build our own WiCi library and replace the original CUDA library at runtime, recording the functions and parameters to send to the server. Two programming interfaces are provided by CUDA: runtime and driver APIs. We choose the driver API as the target for interception because all other CUDA libraries, including the CUDA runtime API, CuDNN, and CUTLASS, are ultimately implemented using CUDA driver APIs. Thus, we can achieve great scalability by selecting this level.

For most CUDA library APIs, we only need to implement the function stub in our library and add our forwarding code. However, kernel functions necessitate special implementation. These functions are vital in GPU-accelerated AI computing, as they are compiled into fatbins and packaged into host-executable binaries. Invoked by the host and executed on GPUs, they manipulate data in VRAM and are crucial for data-parallel processing tasks, such as vector addition and matrix multiplication. Their implementation varies according to the application's needs and the algorithm's design. When executing locally, kernel functions and their inputs are represented as addresses, with input parameters defined by predetermined types. For remote execution, a Fatbin parser is used to parse the function name and input parameters, explicitly obtaining their exact values. This ensures that the correct number and length of inputs are transferred to the remote GPU.

\begin{figure}
    \centering
    \captionsetup{font=small}
    \includegraphics[width=\linewidth]{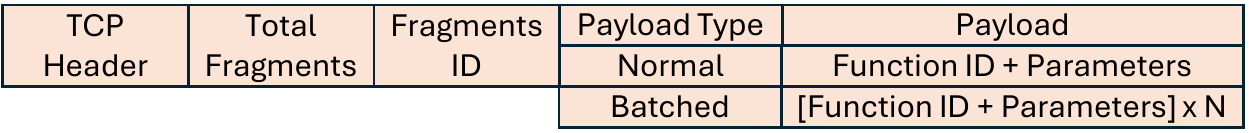}
    \setlength{\abovecaptionskip}{-4mm}
    \caption{Header of WiCi packets.}
    \vspace{-1cm}
    \label{header}
\end{figure}
\noindent\textbf{Forwarding protocol.}
We develop our forwarding protocol based on TCP, and the application-layer packet header is shown in \cref{header}.
Due to the exceedingly large parameter size of certain functions -- with individual functions exceeding 500MB during model loading -- we must first implement fragmentation on the client side.
Fragment-related information occupies two bytes, representing the total number of fragments and the ID of the fragments carried by the current packet.
Next comes the payload type. 
We primarily have two types of payload: normal and batched.
The normal payload type indicates that this remote call invokes only one function. 
The batched type represents a function batch, where the payload contains multiple function IDs and their inputs (\cref{sec:func_batch}). 
After reassembling fragments, the server reads and executes functions based on these flags.

\noindent\textbf{Context management.} We don't need to explicitly manage the CUDA context because we only forward API calls (functions and parameters), with each application corresponding to a thread on the server. This means we are using the native CUDA context on the server, which is managed by the GPU. The advantage of this design is that we can naturally support multiple clients; we just need to start a new server thread to represent each one. And we benefit from the resource isolation provided by the native CUDA context. Further, if needed, we can add a scheduler on the server side to allocate resources for each client (\cref{sec:discussion}).


\vspace{-0.3cm}
\subsection{Challenges}
\label{sec:cha}
In this part, we will introduce the challenges we encountered during the design of our optimization.

\noindent\textbf{Massive data copies.} 
Since we retain all user-space data on the mobile device, including model tensors and function inputs, this data must be forwarded via WiFi to the GPU router. 
Model sizes can reach tens of gigabytes, while function inputs during runtime may require a data rate of tens of megabytes per second. Considering the Falcon 40B model, it takes approximately 85 seconds to load the model locally. However, the time cost associated with WiFi network loading is around 1380 seconds. These massive data copies should ideally be handled by interfaces like PCIe. 
Replacing them with WiFi significantly increases both data transmission latency and base RTT, making the management of these massive data copies one of the major challenges.

\begin{figure*}[t]
     \centering
     \captionsetup{font=small}
     \begin{subfigure}[t]{0.29\linewidth}
            \centering
            \setlength{\abovecaptionskip}{0mm}
            \includegraphics[width=\textwidth]{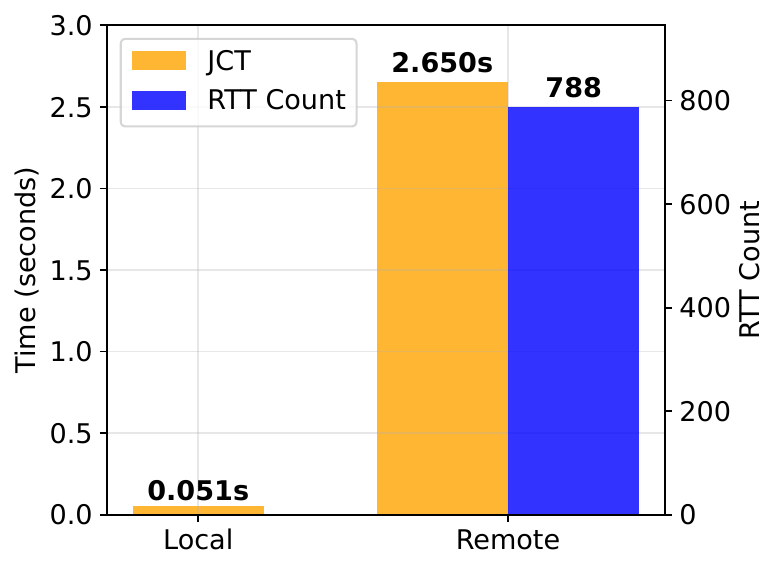}  
            \caption{JCT and RTT count of "nvidia-smi" when executed locally and remotely through kernel driver hijacking.}
            \label{fig:rtt_analysis_chart}  
     \end{subfigure}
     \hspace{-1mm}
     \begin{subfigure}[t]{0.24\linewidth}
            \centering
            \setlength{\abovecaptionskip}{0mm}
            \includegraphics[width=\textwidth]{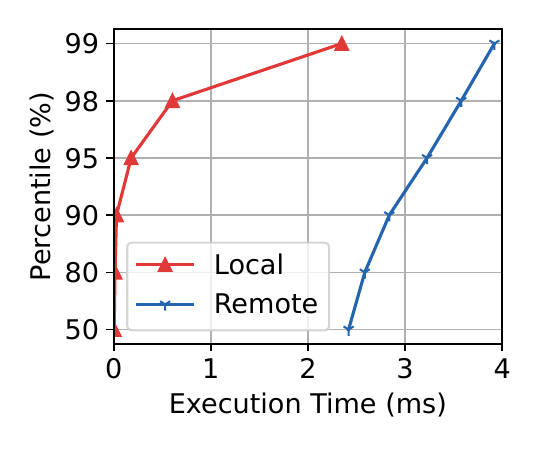}
            \caption{CDF of function execution time. Remote execution can take 100x more time than local execution.}
            \label{fig:maintest_cdf}  
     \end{subfigure}
     \hspace{0mm}
     \begin{subfigure}[t]{0.23\linewidth}
            \centering
            \setlength{\abovecaptionskip}{0mm}
            \includegraphics[width=\textwidth]{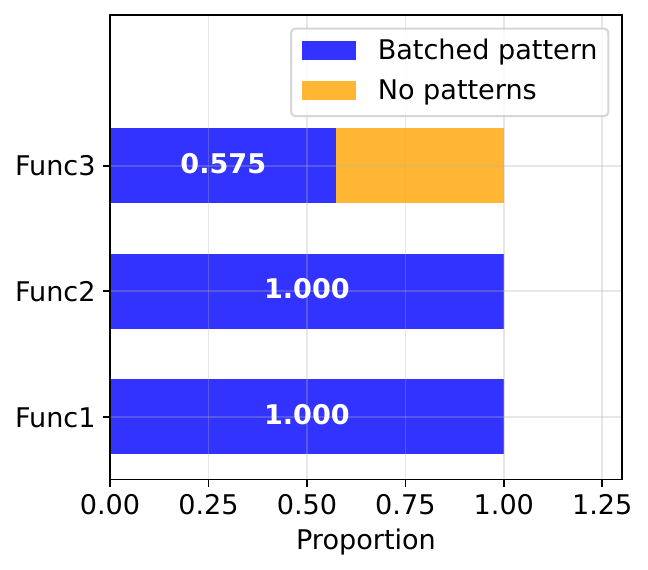}  
            \caption{Proportion of functions invoked under patterns. Many functions can find a pattern.}
            \label{fig:batching}
     \end{subfigure}
     \hspace{1mm}
     \begin{subfigure}[t]{0.15\linewidth}
            \centering
            \setlength{\abovecaptionskip}{0mm}
            \includegraphics[width=\textwidth]{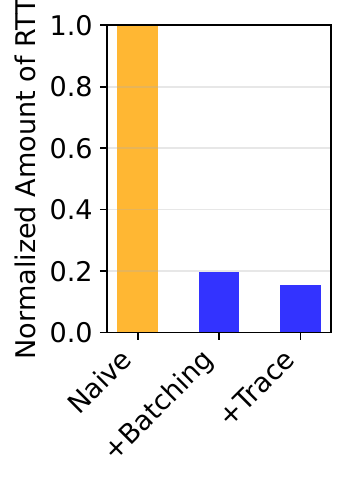}  
            \caption{RTT consumption after optimization.}
            \label{fig:rtt_consum}
     \end{subfigure}
     \setlength{\abovecaptionskip}{0mm}
     \caption{Motivation analysis and evaluation results for function batching and trace replay.}
     \vspace{-4mm}
\end{figure*}

\noindent\textbf{Numerous function calls and prolonged wireless link RTT.}
Through our experiments, running an LLM application locally with a standalone GPU calls approximately hundreds to thousands of functions per second at the NVIDIA driver layer based on the inference speed.
In terms of the estimated time to generate a token by WiCi ($T_{WiCi}$), this can be calculated directly: 
\vspace{-4mm}
\begin{equation}
    T_{WiCi}=T_{Local}+N*RTT+\frac{D }{ Rate}
    \vspace{-1mm}
\end{equation}
$T_{Local}$ denotes the time to generate a token locally by the standalone GPU. $N$ and $RTT$ denote the number of functions called at the NVIDIA driver and base RTT correspondingly. $D$ and $Rate$ denote the Data volume to be transmitted and the data rate.

Apparently, numerous function calls and prolonged wireless link RTT will significantly increase the time taken for inference, being the second challenge of our design.
As an example, we employ the multimodal model Qwen3-VL-8B to perform an image understanding task in one of our experiments, producing outputs of approximately 180 tokens in length. 
Local inference takes less than two seconds. 
Excluding the model loading phase, the whole execution requires roughly 20,000 CUDA user-mode driver function calls, which results in the execution taking over $20\times$ longer JCT due to millisecond-level WiFi RTT.

\section{Optimizations}
To address the challenges we face, we propose threefold optimizations to enhance performance.

\vspace{-0.2cm}
\subsection{Model Caching}
\noindent\textbf{Loading model remotely costs unreasonable time.} In a typical scenario, the application loads the model directly from disk to GPU memory via a PCIe link. In contrast, our system requires the application to first transmit the model from the client to the server over a WiFi link before loading it into GPU memory. Given the significant disparity in bandwidth, the model load time is predominantly limited by the WiFi link. If the model load time is extremely high, it would be a disadvantage of our system, as it can adversely affect the user experience. Every user is eager to load the model as quickly as possible.

\begin{figure}[t]
     \centering
     \captionsetup{font=small}
     \begin{subfigure}[t]{0.48\linewidth}
            \centering
            \setlength{\abovecaptionskip}{0mm}
            \includegraphics[width=\textwidth]{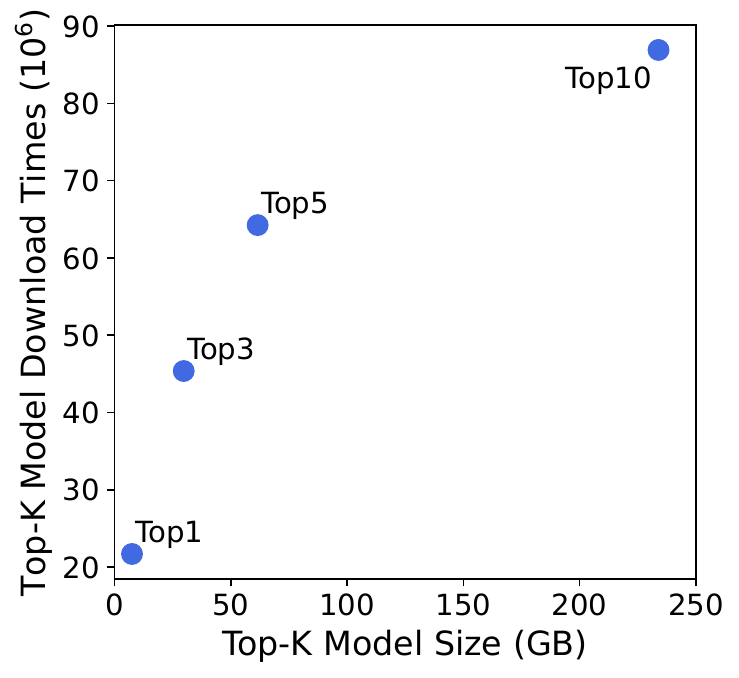}
            \caption{The top 10 models on Hugging Face have been downloaded 90 million times, using only 250 GB.}
            \label{fig:model_caching_hugging_face}  
     \end{subfigure}
     \hspace{1mm}
     \begin{subfigure}[t]{0.48\linewidth}
            \centering
            \setlength{\abovecaptionskip}{0mm}
            \includegraphics[width=\textwidth]{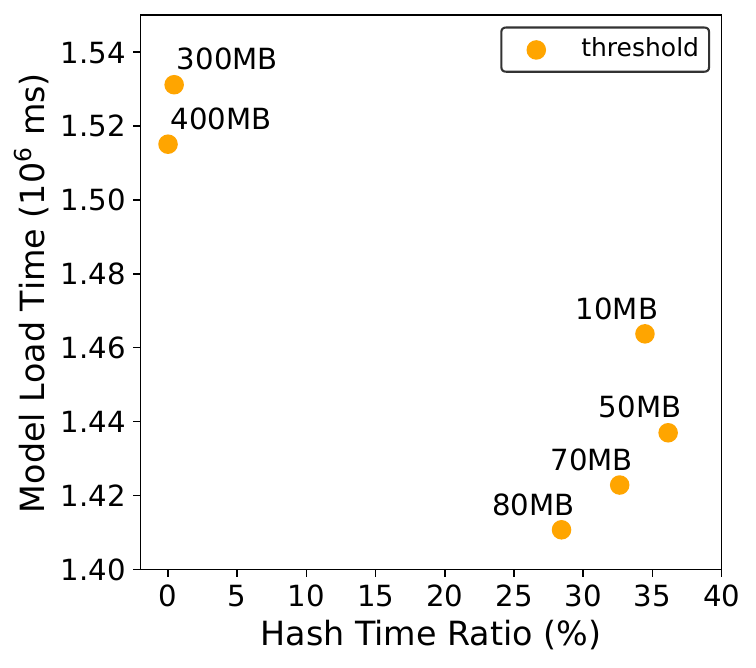}  
            \caption{The initialization model load time for the Falcon 40B varies with different thresholds.}
            \label{fig:model_caching_md5}  
     \end{subfigure}
     \label{fig:model_caching}
     \setlength{\abovecaptionskip}{0mm}
     \caption{Model caching.}
     \vspace{-6mm}
\end{figure}

\noindent\textbf{The popularity of various models exhibits a Matthew effect.} Our observations indicate that users typically engage with only a select number of popular models during a given period. We sort the models on Hugging Face \cite{huggingface} by download times, focusing on those larger than 3 billion parameters. We select the top 10 models, including popular ones like Qwen \cite{qwen} and others with sizes up to 32 billion parameters. \cref{fig:model_caching_hugging_face} displays the total download times and model sizes for these top K models. The data shows download times reaching approximately 90 million, with model sizes around 250 GB, indicating that we can cache popular models within reasonable disk space.

\noindent\textbf{Caching models to speed up loading.} Based on this observation, we propose a design to cache models on the server's disk and retrieve them as needed, thereby eliminating the need for network transmission and saving time. This design is deployed with the API forwarding proxy. The parameters of LLMs are organized and stored as hundreds of tensors in files, which the LLM inference frameworks load tensor by tensor. In order to identify each tensor, we compute the hash value (specifically, the MD5 value in this study; other hash algorithms can also work) on the client side and transmit it to the server. The server then searches for the file named by the hash value. If the server finds it, it sends back a success message. If the file cannot be found, the server sends a request message, and the client then sends the whole tensor, which the server saves in a file. With this model caching optimization, the client only transmits the tensor in the event of a cache miss.

Model caching involves certain costs, notably the time required for clients to compute the hash value, which is significant due to the large size of the model. For smaller tensors, it may be inefficient to calculate and cache them, given that the transmission time cost is relatively low. To address this issue, we establish a threshold that guides the caching decision, allowing us to cache a tensor only when its size exceeds this threshold. \cref{fig:model_caching_md5} illustrates the initial model load time across various thresholds. Our analysis shows that the hash computation time decreases as the threshold increases, which aligns with expectations. Additionally, the model load time initially decreases but then increases as the threshold decreases. This observation is consistent with our hypothesis that the cost of hash computation may exceed the transmission time for smaller blocks.

\vspace{-0.2cm}
\subsection{Batching and Trace Replay}
\label{sec:batching_trace}
\noindent\textbf{Increased latency caused by WiFi networks and enormous function calls.}
Our analysis has ruled out kernel-level function hijacking and forwarding due to poor performance, but user-mode driver implementations also face considerable performance challenges.
Although fewer than kernel drivers, completing a single LLM inference still requires tens of thousands or more user-mode driver calls.
Each individual function requires one RTT to complete if we naively execute them remotely. 
Moreover, many of these functions involve input sizes of tens of thousands of bytes, which further increases latency because the bandwidth of WiFi is much lower than PCIe.
As shown in \cref{fig:maintest_cdf}, we measure the execution time of functions within the user-mode driver during both local execution and remote execution using WiCi of LLM inference, and plot the cumulative distribution function (CDF) graph.
At the median and 90th percentile, remote function execution takes 483 times longer than local execution and 97 times longer, respectively.

\noindent\textbf{Reducing the RTT required for execution is the primary optimization focus.}
As shown in \cref{fig:maintest_cdf}, the gap in function execution latency widens as it approaches the median, while the gap at the same percentile remains nearly unchanged. 
This demonstrates that the baseline RTT of the wireless link is the primary cause of additional latency. 
Generally, the remote execution of a single function introduces one additional RTT. 
To address this issue, the most direct approach is to reduce RTT consumption — in other words, execute multiple functions within a single RTT.
The basic idea is quite straightforward, with two approaches: sending multiple functions together in a batch and skipping certain functions. 
We will now explain how we combined these two concepts into our design.

\vspace{-2mm}
\subsubsection{Function Batching}
\label{sec:func_batch}
During our development and experimental analysis, we observed that many functions occur in fixed patterns during inference. 
That is, certain functions are always called sequentially in a specific pattern.
We further conducted dependency analysis and found that such a patterned method can be categorized into two types:
\begin{itemize}
    \item Outcome-dependent function pattern. As described, this pattern represents a sequence of functions where the later function depends on the execution result of the prior one, such as operations on the same pointer or where the output of the former serves as input for the latter. For LLM inference, typical examples include: "cuMemCpyHToDAsync\_v2 - cuStreamSynchronize" and "cuLibraryLoadData - cuCtxPushCurrent\_v2" (corresponding to Func1 and Func2 in \cref{fig:batching}).
    \item Independent function pattern. These function patterns do not have any sequential dependencies on each other. They simply tend to appear together when invoked, such as "cuLaunchKernel-cuLaunchKernel" (corresponding to Func3 in \cref{fig:batching}).
\end{itemize}
We further calculate how many times the mentioned function is called and how many times it is called by the above three patterns during LLM inference.
As shown in \cref{fig:batching}, the proportion of two dependent patterns reaches one hundred percent. 
The proportion of Func3 reaches 57.5\%.

Based on our findings, we propose function batching that sends several functions together as a batch in one RTT if they match the pattern we set in advance.
If any of the first functions in the patterns are called, we do not forward them to the GPU server directly but instead return a successful execution result locally to obtain the subsequent function.
If the following function calls match the pattern, they are batched and sent together. 
If no match is found, we fall back and resume normal execution.
By doing so, we can execute several functions that originally required multiple RTTs to complete in a single RTT.

\vspace{-3mm}
\subsubsection{Trace replay.} 
Our third optimization is trace replay.
Besides the functions that can be combined into batches, there are still numerous functions that can not be batched by patterns and need to be executed individually. 
However, in practical applications, we have further discovered that patterns exist not only in the sequences between two or three individual functions but also in some function groups comprising dozens of functions.
These function groups share an identical function execution order.
This is actually quite intuitive. 
Take large model inference as an example: the model itself remains unchanged during inference, performing the same computation steps each time it generates a token, with only the input varying. 
Which means, for example, if generating a single token requires thirty functions, their execution order is fixed and repeatedly applied.
Furthermore, the inputs can be derived from those of the previous group, as inputs are directly determined by the token index.
The same principle applies to other applications, such as CNN training.

So here we propose trace replay: we will first record a trace of these repeated function groups, and replay the trace without forwarding any functions to the GPU router.
By doing so, we are "skipping" lots of functions and saving a huge amount of RTTs needed.
We first identify some initial functions that may lead to function groups, which will be referred to as traces in the subsequent content.
Whenever we encounter these initial functions, we begin recording a trace, capturing all subsequent function calls until the next initial function is detected. 
This forms a complete trace and is recorded on both the GPU router and the mobile device.
Next, we will not forward functions directly. 
On mobile devices, we will return success immediately for called functions while verifying whether the function and input match the trace. 
When a trace is fully repeated, we send only the initial function and its parameters, indicating that this trace has been replayed. 
When the trace does not match, we revert to the normal one-by-one execution logic. 
This approach enables us to complete a group of functions with just one RTT and one function's input, significantly reducing the RTT required and data transmission overhead.

After implementing these two optimizations, we calculated the amount of RTT that could be saved. 
The tested application is the same image understanding LLM inference task mentioned in \cref{sec:cha} with an output length of approximately 180 tokens. 
As shown in \cref{fig:rtt_consum}, function batching could reduce RTT consumption by 80.4\% compared to the naive WiCi without any optimizations, while trace replay could further lower RTT consumption by 20.4\%.

\section{Evaluation}
\label{sec:eval}
This section outlines the evaluation setup (§\ref{sec:experiment_setup}), assesses the overall performance of the system against baselines under various conditions (§\ref{sec:overall_performance}), analyzes system capabilities at different request and data rates (§\ref{sec:capability}), and conducts extensive testing of the system using microbenchmarks (§\ref{sec:microbenchmark}).

\vspace{-0.2cm}
\subsection{Experiment Setup}
\label{sec:experiment_setup}
\noindent\textbf{Testbed.} We deploy WiCi on our testbed, which consists of a server equipped with an Intel i7-14700K CPU, 32 GB of memory, and a 4090D 48G GPU. Additionally, we have two client devices, each using a Raspberry Pi 5 with 16 GB of RAM and a 128 GB SD card, both fitted with an Intel AX200NGW network adapter. The server is connected to the router via a wired connection, while the clients connect to the router wirelessly through WiFi. The network connection between the server and the clients is established using TCP.

\noindent\textbf{Baselines.} We have two main baselines: running the model locally on the server, referred to as the server's performance, and running it locally on the client, referred to as the client's local performance. Our goal is to achieve the client's performance with the remote GPU that matches the performance of the model running locally on the server.

\noindent\textbf{Models.} We evaluate WiCi using a series of popular models based on Qwen3 \cite{yang2025qwen3} and a large model derived from Falcon \cite{almazrouei2023Falcon}, which can fully utilize the GPU memory of our server, as detailed in Table \ref{tab:model_conf}. Due to limited memory, we can only run small models on the client device. The largest model we can use is Qwen-8B-GGUF, which helps test the device's performance limits.

\noindent\textbf{Datasets.} We evaluate WiCi using a diverse range of datasets, including MMLU \cite{hendrycks2020measuring}, GSM8K \cite{cobbe2021training}, and Alpaca \cite{taori2023stanford}, as detailed in Table \ref{tab:dataset_conf}. These datasets encompass a variety of linguistic structures, allowing for a comprehensive evaluation across different sparsity patterns. For each dataset, we randomly select 100 samples and evaluate them only on the test set, since profiling or fine-tuning is not needed.

\noindent\textbf{Metrics.} The key metrics for LLM systems are model load time, token rate, and time to first token (TTFT). We measure performance using normalized metrics. And we normalize the results in the following way: all other (client's local and WiCi) performances should be divided by the server's performance, treating the server's performance as 1.

\noindent\textbf{Users. } Apart from the section addressing multiple clients, our primary focus is on examining the system's performance with a single client.

\begin{table}[t]
    \centering
    \captionsetup{font=small}
    \caption{Model configurations. Local means the model is executable on our mobile client, and WiCi means it is executable using WiCi framework.}
    \vspace{-4mm}
    \begin{tabular*}{0.47\textwidth}{@{\extracolsep{\fill}}ccc}
        \toprule
        Model & Size & Client Support \\
        \midrule
        Qwen-0.6B-GGUF   & 639 MB   & local and WiCi   \\
        Qwen-1.7B-GGUF   & 1.83 GB   & local and WiCi   \\
        Qwen-4B-GGUF   & 4.28 GB   & local and WiCi   \\
        Qwen-8B-GGUF   & 8.71 GB   & local and WiCi   \\
        \midrule
        Qwen-14B-GGUF   & 15.7 GB   & WiCi only   \\
        Qwen-32B-GGUF   & 34.8 GB   & WiCi only   \\
        Falcon-40B-GGUF   & 44.5 GB   & WiCi only \\
        \bottomrule
    \end{tabular*}
    \vspace{-4mm}
    \label{tab:model_conf}
\end{table}

\begin{table}[t]
    \centering
    \captionsetup{font=small}
    \caption{Dataset configurations}
    \vspace{-4mm}
    \begin{tabular*}{0.47\textwidth}{@{\extracolsep{\fill}}cc}
        \toprule
        Dataset & Description \\
        \midrule
        MMLU   & Multitask knowledge evaluation benchmark   \\
        GSM8K   & Grade school math problem dataset    \\
        Alpaca   & Instruction-tuning dataset for Llama    \\
        \bottomrule
    \end{tabular*}
    \vspace{-6mm}
    \label{tab:dataset_conf}
\end{table}

\begin{figure*}[t]
     \centering
     \captionsetup{font=small}
     \begin{subfigure}[t]{0.48\linewidth}
            \centering
            \setlength{\abovecaptionskip}{0mm}
            \includegraphics[width=1\textwidth]{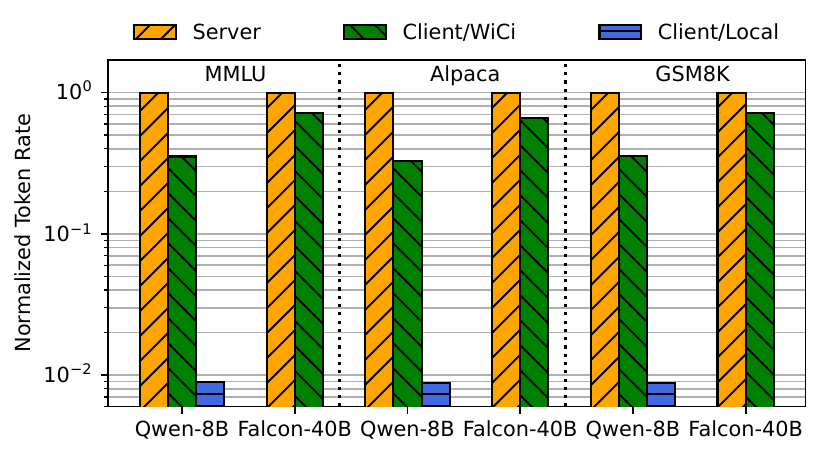} 
            \caption{Normalized Token Rate}
            \label{fig:overall_performance_token_rate}
     \end{subfigure}
     \hspace{1mm}
     \begin{subfigure}[t]{0.48\linewidth}
            \centering
            \setlength{\abovecaptionskip}{0mm}
            \includegraphics[width=1\textwidth]{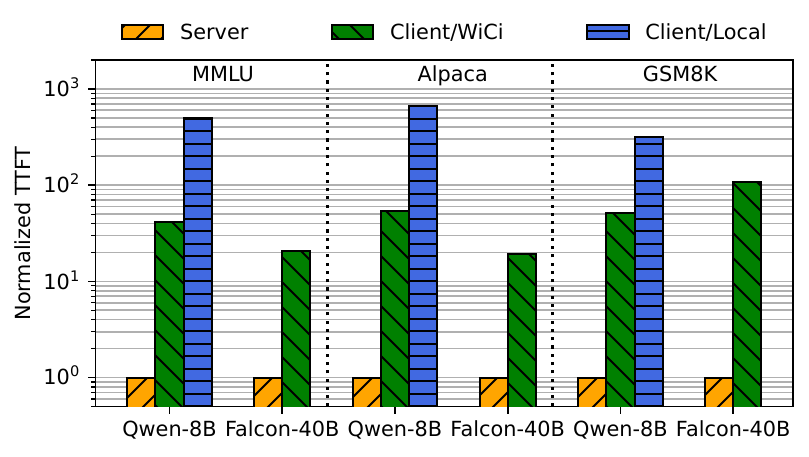} 
            \caption{Normalized TTFT Performance}
            \label{fig:overall_performance_ttft}
     \end{subfigure}
     \setlength{\abovecaptionskip}{0mm}
    \caption{Overall performance is measured by two main factors: (a) normalized token rate, improved by 39x by WiCi, and (b) normalized TTFT, reduced by 90\% by WiCi, evaluated across different LLMs and datasets.}
     \vspace{-4mm}
\end{figure*}

\begin{figure*}[t]
     \centering
     \captionsetup{font=small}
     \begin{subfigure}[t]{0.28\linewidth}
            \centering
            \setlength{\abovecaptionskip}{0mm}
            \includegraphics[width=\textwidth]{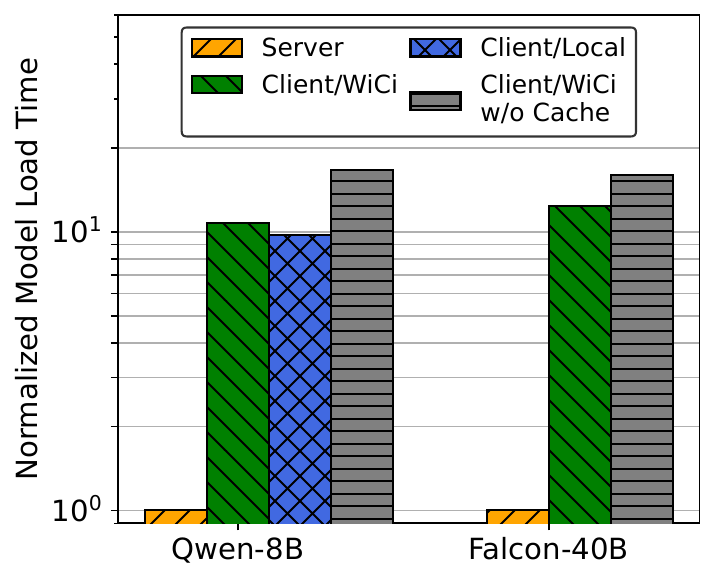}  
            \caption{Model Load Time}
            \label{fig:overall_performance_model_load_time}  
     \end{subfigure}
     \hspace{1mm}
     \begin{subfigure}[t]{0.28\linewidth}
            \centering
            \setlength{\abovecaptionskip}{0mm}
            \includegraphics[width=\textwidth]{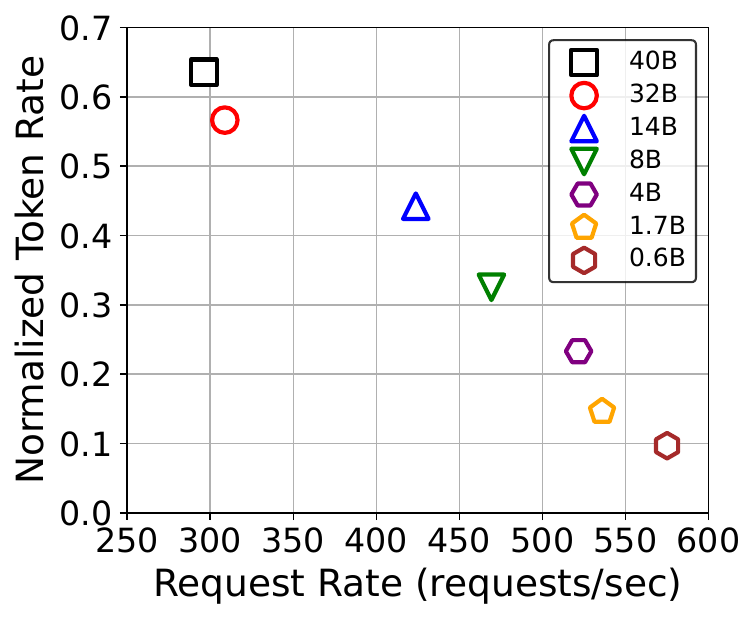}
            \caption{Token Rate-Request Rate}
            \label{fig:capability_request_rate}  
     \end{subfigure}
     \hspace{1mm}
     \begin{subfigure}[t]{0.28\linewidth}
            \centering
            \setlength{\abovecaptionskip}{0mm}
            \includegraphics[width=\textwidth]{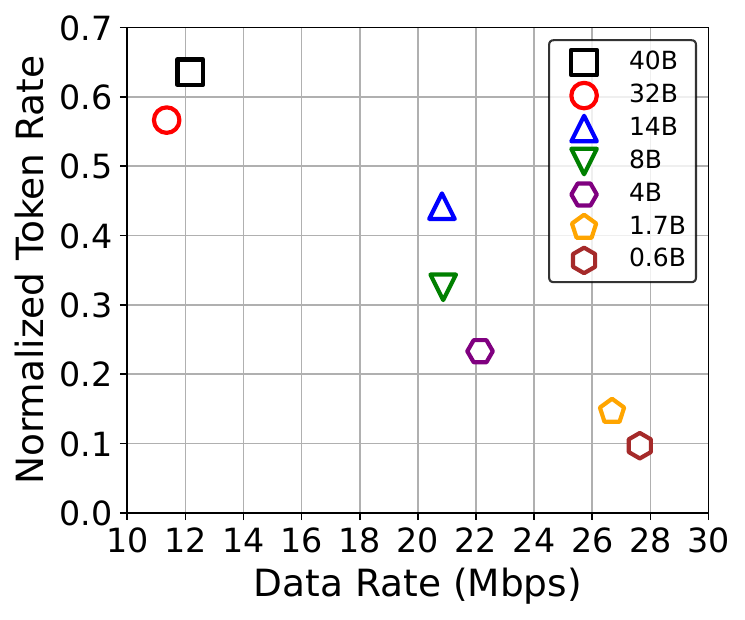}  
            \caption{Token Rate-Data Rate}
            \label{fig:capability_data_rate}
     \end{subfigure}
     \setlength{\abovecaptionskip}{0mm}
     \caption{(a) illustrates the model load time reduced by 36\% by WiCi across different scenarios. (b) and (c) analyze the system's bottlenecks concerning request rate and data rate, respectively.}
     \vspace{-4mm}
\end{figure*}

\subsection{Overall Performance}
\label{sec:overall_performance}
In this section, we present the overall performance of our system WiCi by discussing three key metrics: normalized token rate, normalized TTFT, and normalized model load time. These metrics are essential for evaluating LLM services.

\noindent\textbf{Normalized token rate.} \cref{fig:overall_performance_token_rate} shows the normalized token rate for different LLMs, including Qwen-8B and Falcon, across datasets like MMLU, Alpaca, and GSM8K. The normalized token rates remain stable across these datasets, indicating that WiCi's performance is consistent, as expected. For Qwen-8B, the client's remote normalized token rates are 0.353, 0.326, and 0.355, while the client's local normalized token rates are much lower at 0.0089, 0.0088, and 0.0088. This indicates that using WiCi to enable LLM service can accelerate the token rate by approximately 39$\times$. For Falcon-40B, which cannot be run locally on the client device, the client's remote normalized token rates are 0.713, 0.660, and 0.716. These rates are considered acceptable compared to the scenario where the model cannot be used at all.

\noindent\textbf{Normalized TTFT.} \cref{fig:overall_performance_ttft} employs the same datasets and LLM settings as \cref{fig:overall_performance_token_rate}. TTFT differs from token rate; while token rate is more stable because it averages the time for many tokens, TTFT is more influenced by specific prompts (e.g., datasets). More sophisticated tasks may lead to a higher TTFT, which makes results across different datasets not directly comparable. For Qwen-8B, the WiCi's normalized TTFT values are 41.444, 54.381, and 51.311 on each dataset, respectively, whereas the client's local normalized TTFT values are significantly higher at 501.990, 669.684, and 314.545. This indicates that utilizing WiCi can reduce TTFT by up to 90\%. For Falcon-40B, the WiCi's remote normalized token rates are 20.558, 19.068, and 106.858.

\noindent\textbf{Normalized model load time.} \cref{fig:overall_performance_model_load_time} presents the normalized model load time across various scenarios, along with the performance breakdown of our model caching algorithm. For Qwen-8B, the WiCi's normalized model load time is 10.746, while the client's local normalized model load time is 9.763, indicating that the client's local and remote model load times are nearly the same. Additionally, the WiCi's normalized model load time is 16.646. This shows that our model caching algorithm reduces the model load time by up to 36\%. For Falcon-40B, the WiCi's normalized model load times are 12.365 and 16.011 with and without the model caching algorithm, respectively. In this scenario, the model load time is reduced by 23\%.

\vspace{-0.2cm}
\subsection{Capability}
\label{sec:capability}
The primary distinction between the remote GPU offered by WiCi and a local GPU lies in the data link. Local GPUs utilize a PCIe link for data exchange, which provides lower latency and higher bandwidth compared to the WiFi connection employed by WiCi. This difference results in higher latency and reduced bandwidth for WiCi. Given that LLMs necessitate low latency and high throughput for optimal performance, we evaluate WiCi's performance across various request and data rates in this section to elucidate its capabilities.

\noindent\textbf{Request rate.} \cref{fig:capability_request_rate} presents the normalized token rate performance for various request rates. The request rate is determined by dividing the total number of requests by the runtime. In the graph, the x-value of each data point corresponds to the request rate associated with the respective model. For instance, when deploying the Qwen-0.6B model on WiCi, the anticipated request rate is approximately 575. The normalized token rates corresponding to the respective request rates are as follows: 63.58\% at 296 requests per second, 56.66\% at 208 requests per second, 44.24\% at 423 requests per second, 32.53\% at 469 requests per second, 23.31\% at 521 requests per second, 14.61\% at 535 requests per second, and 9.68\% at 575 requests per second. Overall, the disparity between local and remote performance becomes increasingly pronounced as the request rate decreases. This observation indicates that RTT is a critical factor influencing the system. Additionally, the performance of our system is related to model size, suggesting that if we utilize an advanced GPU, such as the NVIDIA H200 equipped with 144 GB memory, the normalized token rate could approach 1.

\noindent\textbf{Data rate.} \cref{fig:capability_data_rate} illustrates the normalized token rate performance across various data rates. The data rate is calculated by dividing the total amount of packet payload by the runtime. In the graph, the x-value of each data point corresponds to the data rate associated with the respective model. For example, when deploying the Qwen-40B model on WiCi, the anticipated data rate is 12.168 Mbps. The normalized token rates corresponding to the data rates are as follows: 63.58\% at 12.168 Mbps, 56.66\% at 11.356 Mbps, 44.24\% at 20.829 Mbps, 32.53\% at 20.870 Mbps, 23.31\% at 22.139 Mbps, 14.61\% at 26.680 Mbps, and 9.68\% at 27.636 Mbps. Overall, the disparity between local and remote performance becomes increasingly pronounced as the data rate decreases. This observation indicates that bandwidth is a critical factor influencing system performance. It suggests that the availability of advanced WiFi technology, such as WiFi 8, could significantly enhance the capabilities of our system.

\begin{figure}[t]
    \centering
    \captionsetup{font=small}
    \setlength{\abovecaptionskip}{0mm}
    \includegraphics[width=0.48\textwidth]{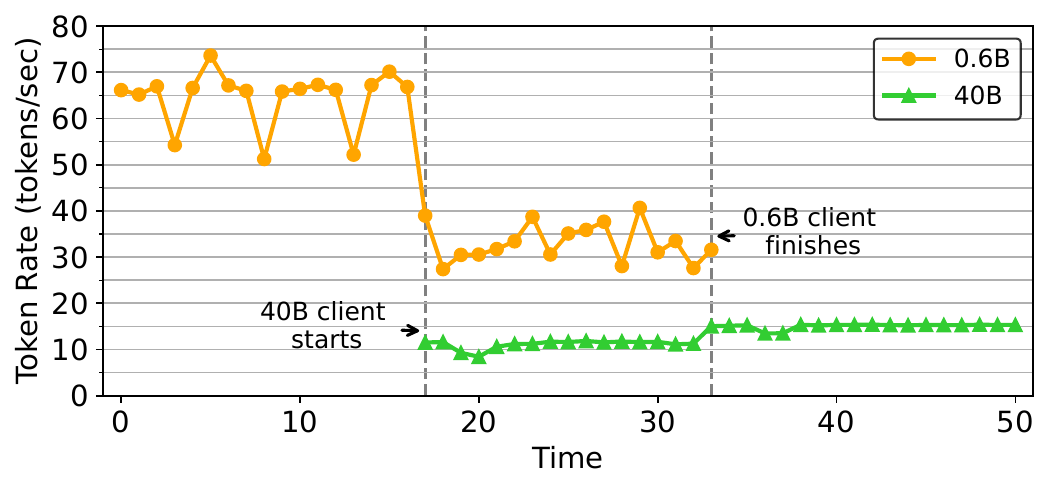} 
    \caption{Client A runs a 0.6B model on the server at first and then client B runs a 40B model later.}
    \label{fig:capability_two_clients}  
     \vspace{-6mm}
\end{figure}

\vspace{-0.2cm}
\subsection{Microbenchmark}
\label{sec:microbenchmark}
\noindent\textbf{Multiple clients.} 
As a wireless computing infrastructure, the system must be capable of supporting multiple clients simultaneously. \cref{fig:capability_two_clients} illustrates the effects on system performance when a new client initiates a connection while an existing client disconnects. Initially, the system establishes a connection with client A using a 0.6B model. The token rate exhibits fluctuations, averaging around 65 tokens per second, which can be attributed to network variability. Subsequently, when a new client, referred to as client B, connects using a 40 B model, the token rate for client A experiences a significant decline due to competition for the data link. Following this, the token rate for client A stabilizes at approximately 35 tokens per second, while client B's token rate averages around 11 tokens per second. After client A disconnects, client B's token rate rises sharply to approximately 15 tokens per second. The results suggest that network conditions may become more complicated in the presence of multiple clients, and we may need to design a congestion control algorithm to mitigate this effect.

\begin{figure}[t]
     \centering
     \captionsetup{font=small}
     \begin{subfigure}[t]{0.48\linewidth}
            \centering
            \setlength{\abovecaptionskip}{0mm}
            \includegraphics[width=\textwidth]{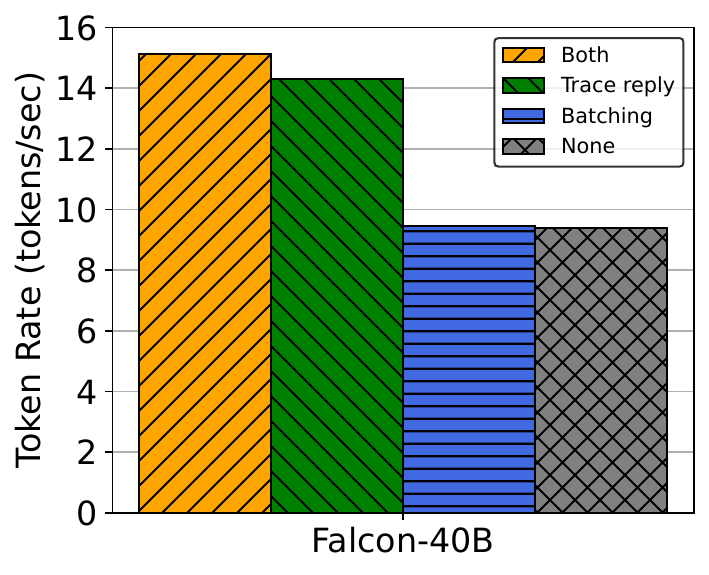}  
            \caption{Token Rate}
            \label{fig:performance_breakdown_token_rate}  
     \end{subfigure}
     \hspace{1mm}
     \begin{subfigure}[t]{0.47\linewidth}
            \centering
            \setlength{\abovecaptionskip}{0mm}
            \includegraphics[width=\textwidth]{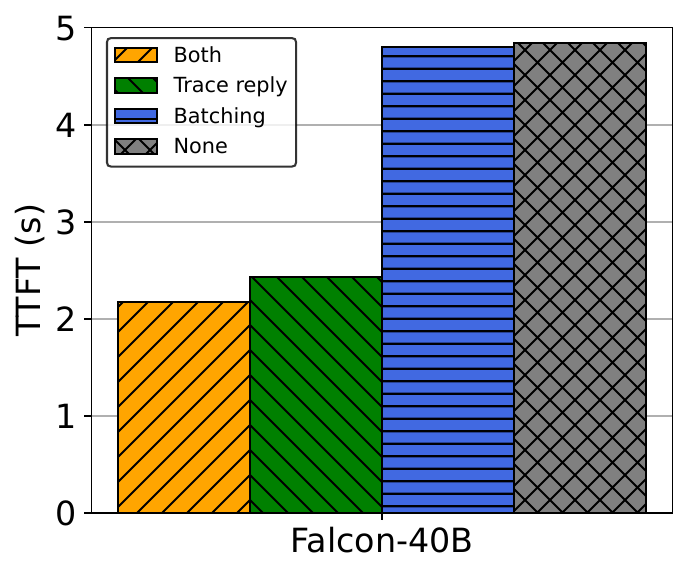}
            \caption{TTFT}
            \label{fig:performance_breakdown_ttft}  
     \end{subfigure}
     \setlength{\abovecaptionskip}{0mm}
     \caption{Both trace reply and batching contribute to the improvement.}
     \label{fig:performance_breakdown}
     \vspace{-6mm}
\end{figure}

\noindent\textbf{Ablation study.} To enhance performance, we propose two designs: trace reply and batching. We conduct an ablation study to evaluate the extent of improvement attributable to each component. \cref{fig:performance_breakdown} illustrates that the implementation of trace reply alone results in a 52.18\% increase in token rate and a 49.82\% reduction in total time to TTFT, demonstrating a significant improvement. In contrast, the improvement achieved solely through batching is modest, with a 0.77\% increase in token rate and a 0.96\% decrease in TTFT. However, when trace reply is combined with batching, the overall improvement is noteworthy, yielding an additional 5.54\% increase in token rate and a further 10.52\% decrease in TTFT.

\begin{figure}[t]
     \centering
     \captionsetup{font=small}
     \begin{subfigure}[t]{0.55\linewidth}
            \centering
            \setlength{\abovecaptionskip}{0mm}
            \includegraphics[width=\textwidth]{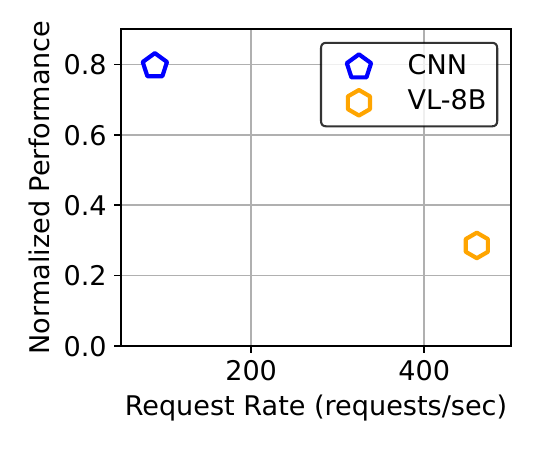}  
            \caption{Normalized performance for\\ diverse applications.}
            \label{fig:multi_app}  
     \end{subfigure}
     \hspace{-4mm}
     \begin{subfigure}[t]{0.47\linewidth}
            \centering
            \setlength{\abovecaptionskip}{0mm}
            \includegraphics[width=\textwidth]{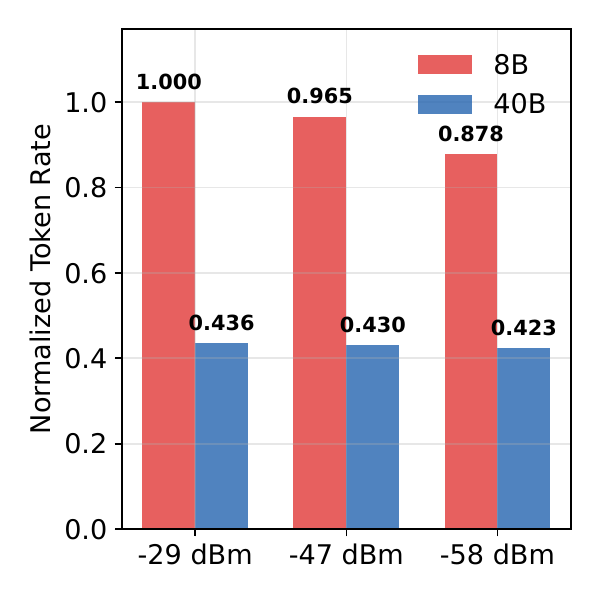}
            \caption{Normalized token rate under varying channel conditions.}
            \label{fig:rssi}  
     \end{subfigure}
     \setlength{\abovecaptionskip}{0mm}
     \caption{Evaluation results for varying applications and channel conditions. WiCi provides robust performance across various tests.}
     \vspace{-6mm}
\end{figure}

\noindent\textbf{Multiple applications.}
Besides LLM inference, WiCi also supports diverse applications with solid performance. 
We select multimodal inference and CNN training as representative applications for evaluation. 
We choose Qwen3-VL-8B for image understanding tasks, and we use NVIDIA's sample for CNN training with an input layer sized 4x1024x224x224, a 3x3 filter, and a stride of 2.
The results are plotted in \cref{fig:multi_app}, the normalized performance here means the inference speed of WiCi compared to the inference speed of the 4090D.
For image understanding tasks, WiCi achieves 28.5\% of the native performance of the 4090D standalone GPU.
For CNNs, WiCi achieves 79.6\% of the native performance.

\noindent\textbf{Varying channel conditions.}
Since we conduct numerous data transmissions over WiFi, a key characteristic of wireless channels is their fluctuating conditions. 
Therefore, we test WiCi's performance of the 8B and 40B models under varying Received Signal Strength Indications (RSSIs).
As shown in \cref{fig:rssi}, at -47 dBm, the inference speeds of the 8B and 40B models decreased by 3.5\% and 1.4\%, respectively. 
At -58 dBm, the inference speeds of the 8B and 40B models decreased by 9\% and 1.6\%, respectively, compared to -47 dBm.
These results demonstrate that WiCi maintains stable performance even under varying channel conditions.

\section{Related Work}
\noindent\textbf{GPU Sharing.} Numerous studies have explored GPU multiplexing for deploying applications in specific scenarios, such as AI training \cite{bai2020pipeswitch, huang2019gpipe, lim2021zico, narayanan2019pipedream, wang2021wavelet, xiao2020antman} and AI inference \cite{cui2021enable, dhakal2020gslice, gu2023fast, gujarati2020serving, shen2019nexus}. These applications are typically deployed in virtual machines or lightweight environments, such as bare-metal servers and containers. GPU sharing across containers or bare-metal environments predominantly utilizes API remoting to manage GPU resources. The host system determines whether and when to pass the intercepted calls. Notable implementations, such as GaiaGPU \cite{gu2018gaiagpu}, qCUDA \cite{lin2019qcuda}, and vCUDA \cite{shi2011vcuda}, utilize API remoting in user space to limit the resource usage of applications that utilize CUDA. These studies primarily focus on sharing GPU resources among users on the same host and do not address the challenges associated with targeting remote mobile devices.

\noindent\textbf{Model quantization and pruning.} Pruning methods \cite{liang2021pruning, liu2020pruning} are designed to reduce the number of parameters in a model while minimizing performance degradation. Several studies \cite{han2015deep, hanson1988comparing, huang2018data, lebedev2016fast, yuan2006model} have investigated static pruning, wherein parameters are pruned offline prior to deployment. In contrast, dynamic sparsity methods \cite{bengio2015conditional, davis2013low, guo2016dynamic, he2018soft, sanh2020movement, wang2025lemo, xu2021rethinking} determine which parameters to prune during runtime, allowing for seamless integration with both training and inference processes.

Model quantization \cite{cheng2017survey, gholami2022survey} seeks to achieve similar objectives by reducing the precision of model parameters, thereby optimizing the utilization of available bits to encode model information more efficiently. A substantial body of research has progressively lowered precision, with efforts encompassing 8-bit representations \cite{dettmers2208int8, xiao2023smoothquant}, 4-bit representations \cite{Bitsandbytes, frantar2022gptq, yao2022zeroquant}, 2-bit representations \cite{chee2023quip}, and even 1-bit representations \cite{wang2023bitnet, xu2024onebit}.

Both of these methods compromise some model performance to reduce computational costs, thereby facilitating the deployment of LLMs on mobile devices. In contrast, WiCi enables the implementation of LLMs on mobile devices without sacrificing performance.

\section{Discussion}
\label{sec:discussion}
\noindent\textbf{Cost for users.}
Our design requires users to purchase their own GPUs with certain hardware specifications.
However, this does not mean it is not cost-effective for users. As mentioned earlier, subscribing to ChatGPT Pro for ten months can cover the cost of an RTX 5090. Purchasing a GPU as a one-time investment avoids costs that would otherwise grow indefinitely over time. Moreover, subscription services only grant access to predefined functionalities within their plans.
Owning your own GPU not only brings you own inference systems with more options and better privacy. It also allows usage across other applications beyond AI inference, such as gaming, video editing, and more, further spreading out the cost across multiple uses.

\noindent\textbf{Task scheduling.}
When multiple clients compete for computing resources, how to schedule their computing tasks to achieve higher performance has long been a critical focus in both academia and industry.
Numerous solutions have been proposed \cite{zhang2025efficient, guardian, nvshare}.
WiCi is actually completely orthogonal to these scheduling algorithms. 
Our function hijacking and forwarding policies inherently distinguish different tasks from different clients on the GPU router side. 
This mirrors the scheduling scenarios in multi-tenant clusters, allowing these algorithms to be directly integrated into WiCi.

\section{Limitation}

\noindent\textbf{Scalability to other GPU applications.}
Beyond AI computing, the WiCi system can also be applied to numerous other GPU applications, such as video encoding and decoding, gaming, and more.
The function calls and underlying patterns in these applications may differ from AI inference. 
Yet, some applications like rendering generate large volumes of data transmission with relatively few function calls compared to AI computing. 
We still need further development and optimization.
We leave expanding the scalability to other GPU applications as our future work.

\noindent\textbf{Competition with Internet traffic.}
We place the GPU on the WiFi router to meet requirements for wireless connectivity, power supply, and other needs. 
However, the continuous data transmission and high bandwidth demands during AI computations — especially during model loading — pose significant challenges to the router's capacity. 
This may also degrade the user experience for other regular internet users sharing the same network. 
To mitigate this, we currently select the idle channel (there are tens of channels at the 5GHz frequency band) with few normal traffic. 
In the future, we will schedule the airtime to ensure that the "bursty" nature of GPU instructions is serviced immediately, and minimize the impact on Internet traffic.


\section{Conclusion}
This paper proposes WiCi, a wireless GPU computing infrastructure that enables mobile devices to wirelessly access a standalone GPU without adding any hardware, as if a truly zero-weight GPU is directly connected to the mobile devices.
WiCi is scalable to applications, compatible with devices, and could provide comparable performance compared to standalone GPUs.

This work does not raise any ethical issues.

\bibliographystyle{ACM-Reference-Format}
\bibliography{reference}

\end{document}